\documentclass[useAMS,usenatbib]{mn2e}
\usepackage{psfig}
\usepackage{epsfig}
\usepackage{references}
\bibpunct{(}{)}{;}{a}{}{,}

\def\dnu{$\Delta\nu$}
\def\dn{$\Delta\nu$}

\def\dnsatd{$\Delta\nu_{\rm sa, 32s}$}
\def\dnsaus{$\Delta\nu_{\rm sa, 1s}$}
\def\dnb{$\overline{\Delta\nu}$}
\def\nudrift{$\overline{|\nu_{\rm drift}|}$}
\def\dnq{$\overline{\rm Q}$}
\def\nuqpo{$\overline{\nu_{\rm qpo}}$}

\begin{document}
\title {On the high coherence of kilo-Hz Quasi-Periodic Oscillations}
    \author[Barret et al.]{D. Barret$^{1}$\thanks{E-mail: Didier.Barret@cesr.fr}, W. Klu\'zniak$^{1,2,3}$, J.F. Olive$^{1}$,
    S. Paltani$^{4}$ and G. K. Skinner$^{1}$\\
$^1$Centre d'Etude Spatiale des Rayonnements, CNRS/UPS, 9 Avenue du Colonel Roche, 31028 Toulouse Cedex 04, France\\
$^2$Institute of Astronomy, Zielona G\'ora University, ul. Lubuska 2, 65-265 Zielona G\'ora, Poland\\
$^3$Copernicus Astronomical Center, ul. Bartycka 18, 00-716 Warszawa, Poland\\
$^4$Laboratoire d'Astrophysique de Marseille, BP 8 - Traverse du Siphon, 13376 Marseille cedex 12, France
}

\date{Accepted 2004 July XX. Received 2004 July XX; in original form 2004 July 23}

\pagerange{\pageref{firstpage}--\pageref{lastpage}} \pubyear{2004}

\maketitle

\label{firstpage}

\begin{abstract}We have carried out a systematic study of the
properties of the kilo-Hertz quasi-periodic oscillations (QPO)
observed in the X-ray emission of the neutron star low-mass X-ray
binary 4U1608-52, using archival data obtained with the Rossi
X-ray Timing Explorer. We have investigated the quality factor, Q, of
the oscillations (defined as the ratio, $\nu / \Delta \nu$, of the
frequency $\nu$ of the QPO peak to its full width at half maximum $\Delta \nu$).
In order to minimise the effect of long-term frequency drifts,
power spectra were computed over the shortest times permitted by
the data statistics. We show that the high Q of $\sim 200$ reported by Berger et
al (1996)  for the lower frequency kilo-Hz QPO in one of their observations is by no means exceptional,
as we observe a mean Q value in excess of 150 in 14 out of the 21 observations analysed and Q can remain above 200 for thousands of seconds. The frequency of the QPO varies over the wide
range 560--890 Hz and we find a systematic trend for the coherence time
of the QPO, estimated as $\tau=Q /(\pi \nu)=1/(\pi\Delta\nu)$, to increase with
$\nu$, up to a maximum level at $\sim 800$ Hz,
beyond which it appears to decrease, at frequencies where the QPO weakens.
There is a more complex relationship
between $\tau$ and the QPO root mean squared amplitude (RMS), in
which positive and negative correlations can be found. A higher-frequency QPO, revealed by correcting for the frequency drift of
the 560--890 Hz one, has a much lower Q ($\sim 10$) which does not
follow the same pattern. We discuss these results in the framework of competing QPO models and show that those involving clumps orbiting within or above the accretion disk are ruled out.
\end{abstract}

\section{Introduction}

Fourier analysis reveals the X-ray emission of low-mass X-ray
binaries (LMXBs) to be variable on short timescales, often with several
characteristic quasi-periodic oscillation (QPO) peaks identified
between $\sim1\,$ Hz and $\sim1\,$ kHz \citep{vdk1,vdk2,vdk3}. The
highest-frequency QPOs, reaching up to $\sim 1300$ Hz and usually
occurring in pairs, are of particular interest, as their periods
correspond to the dynamical time-scale in the inner accretion
disk, where strong-field effects of gravity are crucially
important.  There is no agreement as to the physical origin of the
QPOs \citep{vdk3}. In the wealth of observational papers published on QPOs,
many have studied the QPO properties as a function of count rate,
luminosity, spectra, lower-frequency features and so on, but very
few have focussed on the coherence time, which is in fact a very
constraining parameter for QPO models. In this paper, we study the
previously reported kilo-Hz QPOs in the X-ray burster, 4U 1608-52,
with the aim of characterizing the quality factor, and hence
coherence time, of its QPO over a wide range of frequency. We have
selected this source for two main reasons: first its QPOs are
known to be strong, and second in the QPO discovery paper of
\citet{berger96}, a quality factor $Q=\nu/\Delta\nu$ of up to $\sim 200$ was reported for the QPO, together
with apparent correlations between its frequency and both its
amplitude and width.

\begin{figure*}
\includegraphics[width=0.95\textwidth]{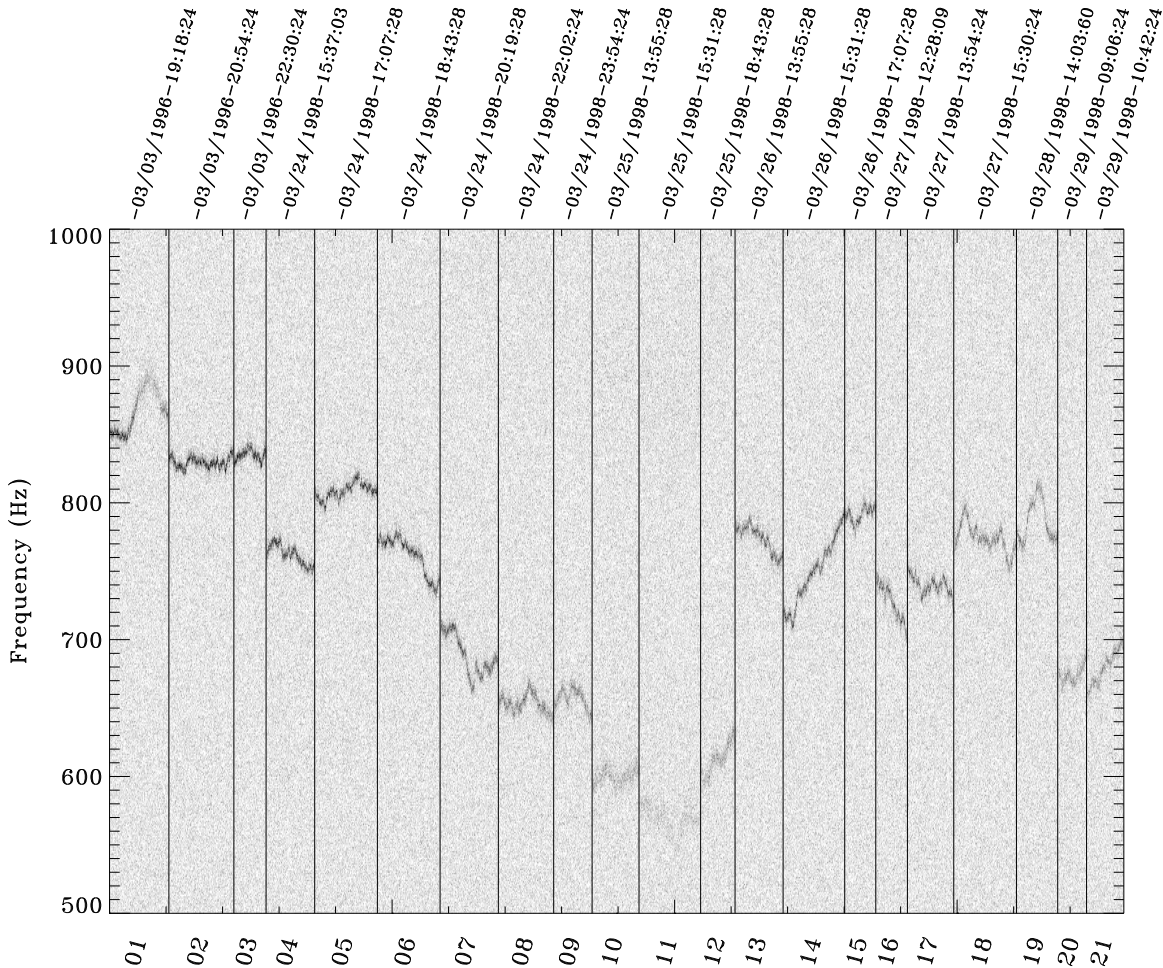}
\vspace*{-1.0cm}\caption[]{Time evolution of the 560-890 Hz lower-frequency QPO detected
from 4U1608-52 over the whole data set. The image is a
dynamical power density spectrum, in which each column is a PDS
integrated over 32 s. The QPO corresponds to the boldest
parts of the image. Note that for the longest segment of 1996
March 25th (segment 11)  the QPO is barely visible in the image.}
\label{DBarret_f1}
\end{figure*}

\section{Data analysis and results}
We have retrieved the 1996 and 1998 data of 4U1608-52 from the
RXTE archive. These data have already been discussed in
\citet{berger96} and \citet{mendez98a,mendez98b,mendez99}. The 1996 data were
recorded during the decline of an X-ray outburst, whereas the 1998
data sampled a whole outburst phase from the source. Here, we have
restricted the analysis to the {\it science event} unbinned data.
We have thus excluded the 1996 March 6th data which were obtained in
a binned data mode with no spectral information. As the
sensitivity for QPO detection increases with the number of photons
detected, we have selected only those observations in which the
five RXTE proportional counter array (PCA) units were operating
\citep{bradt93}. This represents more than 250 kiloseconds of
data. The dataset is divided into 21 segments recorded within
different, sometimes consecutive, orbital revolutions of the satellite.

We have computed Leahy normalized Fourier power density spectra (PDS) between 1 and 2048
Hz over 1 s intervals (1 Hz resolution), using X-ray photons of energy between 2 and 40 keV. It is well known
that the QPO frequency can vary by as much as 50 Hz in a few
thousand seconds. The goal of our analysis is to estimate Q, while
keeping the contribution of the frequency shifts as small as
possible. This implies that the QPO profile properties have to be
estimated on the shortest possible timescales. For 4U1608-52,
which shows strong QPOs, this can usually be done with 32 s.
So we have averaged 32 one-second PDS  and searched for a QPO peak using a
sliding window algorithm, as described in \citet{boirin00}. In
segment 11, where the QPO is the weakest, we have averaged 128 one-second
PDS. We have preferred to use 32 summed 1 Hz spectra rather than a single power spectrum obtained over, say, 32 s as in this way we can make a direct comparison with the results discussed below (\S 2.2), which rely on having 1 s time resolution. Using 32 s spectra would formally give better frequency resolution, but we have found that the reduction in the measured widths and the improvement in the uncertainties in the parameters are very small.

\begin{table*}
\caption{Mean Q factor of the low-frequency kilo-Hz QPO observed from 4U1608-52 in the 21 segments of data.  The count rate is given in cts s$^{-1}$ in the 2-40 keV range for the 5 PCA units. N$_{\rm int}$ is the number of 1s PDS averaged. N$_{\rm s}$ is the number of averaged PDS.  N$_{\rm d}$ is the number of PDS in which the QPO was detected above the 6 $\sigma$ threshold. $\nu_{\rm min, max}$ are the minimum and maximum QPO frequencies measured in the 32 s PDS. The RMS is the root mean square amplitude given as a percentage of the source count rate. \dnb~is the fitted FWHM of the QPO obtained, by shifting and adding the 32 s (128 s for segment 11) PDS to a reference frequency given as $\overline{\nu_{\rm qpo}}$. 
$\overline{Q}$ is obtained by dividing the latter by \dnb.}
\begin{center}
\begin{tabular}{cccccccccccc}
Segment & Date & cts s$^{-1}$ & N$_{\rm int}$ & N$_{\rm s}$  & N$_{\rm d}$  & $\nu_{\rm min, max}$ & RMS (\%) & $\overline{\nu_{\rm qpo}}$ & \dnb & $\overline{Q}$ \\
\hline
\hline
 1 & 03/03/96--19:18 & 3165.4 &  32 & 106 &  81 & 846.0--891.5 &  6.6$\pm$0.1 & 866.6$\pm$0.1 &  5.8$\pm$0.2 & 150.4$\pm$5.0 \\
 2 & 03/03/96--20:54 & 3041.7 &  32 & 115 & 115 & 821.7--840.0 &  7.7$\pm$0.1 & 829.5$\pm$0.0 &  4.1$\pm$0.1 & 200.4$\pm$4.6 \\
 3 & 03/03/96--22:30 & 3079.0 &  32 &  57 &  57 & 827.5--842.3 &  7.7$\pm$0.1 & 834.5$\pm$0.1 &  4.2$\pm$0.1 & 199.7$\pm$6.3 \\
 4 & 03/24/98--15:37 & 1971.9 &  32 &  86 &  86 & 749.0--773.6 &  9.2$\pm$0.1 & 761.5$\pm$0.0 &  4.2$\pm$0.1 & 181.4$\pm$5.1 \\
 5 & 03/24/98--17:07 & 2060.3 &  32 & 111 & 110 & 795.5--820.5 &  8.6$\pm$0.1 & 808.6$\pm$0.0 &  3.8$\pm$0.1 & 212.7$\pm$5.6 \\
 6 & 03/24/98--18:43 & 1959.1 &  32 & 111 & 111 & 732.2--779.2 &  8.9$\pm$0.1 & 761.5$\pm$0.0 &  4.0$\pm$0.1 & 192.7$\pm$5.0 \\
 7 & 03/24/98--20:19 & 1830.4 &  32 & 103 & 103 & 662.5--710.7 &  9.2$\pm$0.1 & 688.5$\pm$0.1 &  4.9$\pm$0.1 & 139.2$\pm$4.0 \\
 8 & 03/24/98--22:02 & 1767.9 &  32 &  98 &  86 & 642.7--669.6 &  8.7$\pm$0.1 & 653.4$\pm$0.1 &  6.0$\pm$0.2 & 108.9$\pm$3.9 \\
 9 & 03/24/98--23:54 & 1425.7 &  32 &  68 &  46 & 640.2--667.1 &  9.0$\pm$0.2 & 657.4$\pm$0.1 &  5.7$\pm$0.3 & 116.3$\pm$5.9 \\
10 & 03/25/98--13:55 & 1564.3 &  32 &  83 &  20 & 591.7--609.0 &  7.8$\pm$0.2 & 600.5$\pm$0.2 &  9.5$\pm$0.6 &  63.5$\pm$4.0 \\
11 & 03/25/98--15:31 & 1543.4 & 128 &  27 &   8 & 560.9--580.5 &  6.5$\pm$0.2 & 570.1$\pm$0.6 & 16.3$\pm$1.7 &  35.0$\pm$3.6 \\
12 & 03/25/98--18:43 & 1529.1 &  32 &  61 &  23 & 597.1--633.2 &  8.3$\pm$0.2 & 611.1$\pm$0.2 &  9.0$\pm$0.6 &  68.0$\pm$4.8 \\
13 & 03/26/98--13:55 & 1716.8 &  32 &  85 &  84 & 754.7--788.2 &  8.9$\pm$0.2 & 773.5$\pm$0.1 &  4.3$\pm$0.2 & 178.6$\pm$6.3 \\
14 & 03/26/98--15:31 & 1586.5 &  32 & 109 & 109 & 709.5--793.4 &  9.7$\pm$0.1 & 749.5$\pm$0.0 &  4.2$\pm$0.1 & 176.9$\pm$5.0 \\
15 & 03/26/98--17:07 & 1655.6 &  32 &  55 &  54 & 776.3--799.1 &  9.3$\pm$0.2 & 791.4$\pm$0.1 &  4.3$\pm$0.2 & 182.2$\pm$7.4 \\
16 & 03/27/98--12:28 & 1082.0 &  32 &  56 &  55 & 697.3--745.5 & 12.0$\pm$0.2 & 725.5$\pm$0.1 &  4.7$\pm$0.2 & 153.2$\pm$6.5 \\
17 & 03/27/98--13:54 & 1066.3 &  32 &  82 &  82 & 720.6--752.6 & 11.6$\pm$0.2 & 738.5$\pm$0.1 &  4.2$\pm$0.1 & 177.5$\pm$6.2 \\
18 & 03/27/98--15:30 & 1132.9 &  32 & 111 &  97 & 751.0--797.0 & 10.5$\pm$0.2 & 773.4$\pm$0.1 &  4.9$\pm$0.2 & 159.4$\pm$5.5 \\
19 & 03/28/98--14:03 &  942.0 &  32 &  73 &  55 & 767.6--813.2 & 11.0$\pm$0.3 & 786.6$\pm$0.1 &  4.3$\pm$0.2 & 183.0$\pm$8.4 \\
20 & 03/29/98--09:06 &  766.3 &  32 &  51 &  31 & 664.7--688.9 & 12.2$\pm$0.4 & 674.6$\pm$0.1 &  5.8$\pm$0.4 & 117.2$\pm$7.4 \\
21 & 03/29/98--10:42 &  759.0 &  32 &  65 &  41 & 661.8--700.0 & 12.1$\pm$0.3 & 679.6$\pm$0.1 &  4.8$\pm$0.3 & 141.6$\pm$8.0 \\\hline
\end{tabular}
\end{center}
\label{DBarret_t1}
\end{table*}

 To obtain reliable fits, only features of significance larger than $6\sigma$ are considered, although decreasing this threshold to $5\sigma$ does not affect the results. In our data set, QPOs are detected on 1996 March 3rd and between 1998 March 24th and 29th  (see Figure \ref{DBarret_f1} and Table \ref{DBarret_t1}). 

Over the short integration times considered here, the effects of statistical fluctuations on the observed QPO profile can be very important (e.g., all the power can be concentrated into a single frequency bin), hence fitting of the QPO has to be done with great care to minimize potential biases. The QPO is fitted over a 100 Hz frequency range (50 Hz on each side of the QPO peak) with a Lorentzian having three parameters (amplitude, frequency and \dn)  plus a constant representing the counting-statistics noise level (generally very close to the theoretical value of 2). We note that the noise level can also be estimated by averaging the power in a region where no signal is expected, e.g. above 1400 Hz. Although the two methods give very similar results, we find that there is a trend for the latter method to give slightly lower $\Delta\nu$. This is because the fitted mean level in the region of the QPO is generally very marginally lower than that above 1400 Hz, probably due to subtle dead time effects (e.g. \cite{zhang95}). As we are interested in this paper in setting upper bounds to the QPO width, we have preferred to fit the constant level together with the QPO.

In a more dramatic way, the results of the fits are sensitive to the way the error bars are set on the measurements of power. In a Leahy normalised PDS, the theoretical error bars on the power are $P/\sqrt N$, where N is the number of PDS averaged. However simply using the power $P$ in each bin to estimate the error on that bin leads to a bias caused by statistical fluctuations, because bins which happen to have low values of $P$ are given high weight. We have therefore used an iterative procedure in which a first fit is done with uniform error bars and subsequent iterations are made in which the $P$ used in estimating the error bars is based on the fitted model. The uncertainties in the fitted parameters are computed such that $\chi^2=\chi^2_{\mathrm{min}}+1$, with the other parameters refitted where $\chi^2_{\mathrm{min}}$ is the $\chi^2$ corresponding to the best-fit value.

\subsection{Mean QPO width}
First we wish to estimate the mean width (\dnb) of the QPO for each of the 21 segments of data, based on the measurements made every 32 s. For this purpose, we apply the shift-and-add technique (e.g., M{\' e}ndez et al. 1998) to the 32 s PDS.  Each 32 s PDS is fitted to obtain the QPO frequency, then shifted to a reference frequency. The resulting PDS is then averaged to produce a single PDS, from which the mean QPO width (\dnb) can be fitted. The reference frequency chosen is the mean QPO frequency for the segment of data (\nuqpo). 

In those 32 s PDS where the QPO is not detected above $6\sigma$, the QPO frequency used for the shifting of the PDS has been estimated by linearly interpolating between neighbouring detected frequencies on either side. Similar results are obtained if we use a spline interpolation. We have also checked that increasing the PDS averaging time (say to 64 seconds) to increase the number of QPO detections yielded consistent results within error bars for \dnb. 

This method should overestimate somewhat the true \dnb\ because the frequencies determined every 32 s and used in the shift-and-add have typical error bars between 0.5 and 1 Hz, introducing a blurring of the QPO profile. Obviously any variation faster than 32 s would also broaden the QPO. The mean quality factor (\dnq), obtained by dividing \nuqpo\ by \dnb~should thus be considered as a lower limit. Nevertheless, we find values of \dnq\ larger than $\sim 150$ in 14 out of the 21 segments analysed, and values as high as 213 are found, indicating that the large Q factor reported by Berger et al. (1996) is by no means exceptional. The results are listed in Table 1.

\begin{figure} 
\includegraphics[width=0.45\textwidth]{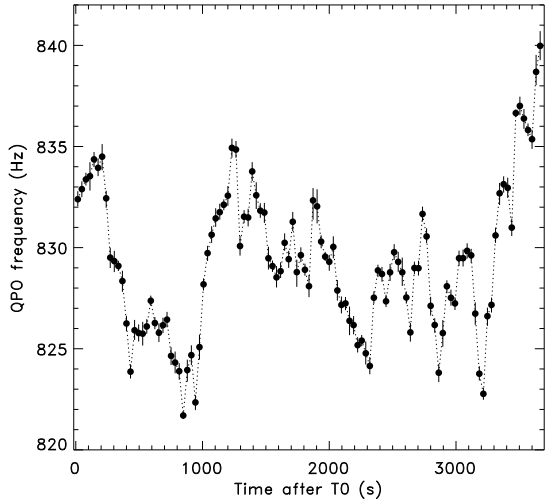}
\caption[]{Time evolution of the QPO frequency in segment 2 as followed every 32 seconds. Frequency drifts of up to 5 Hz in 32 s are observed.}
\label{DBarret_f2}
\end{figure}

\subsection{Attempting to correct for the observed drift}
Even with PDS averaged over only 32 s, the measured width may still be affected by the frequency drift. As can be seen in Figure \ref{DBarret_f2}, drifts as large as 5 Hz between consecutive 32 s PDS are not unusual. Can we estimate what would be the width of the QPO in a one-second PDS (i.e. with negligible contribution from the frequency drift)? It is impossible to fit directly the one-second PDS but one can still make some first order estimates.

If one assumes that the QPO frequency drifts linearly, one can remove the drift contribution using the quadratic relation
\begin{equation}
\overline{\Delta \nu}^2 \sim \overline{\Delta \nu_{\rm 1 s}}^2 + \overline{|\nu_{\rm drift}|}^2
\label{DBarret_eq1}
\end{equation}
as an approximation, where \dnb\ is the mean QPO measured on 32 s with the method described above, $\overline{\Delta \nu_{\rm 1 s}}$ is the mean width expected in a 1 s PDS and \nudrift\ is the mean absolute frequency differences between consecutive 32 s PDS. 

Alternatively, one can try to account for the drift by again using the shift-and-add technique, applied this time to the 1 s PDS. For this, we must estimate the QPO frequency every second. To do so, here we have used a sliding-window technique; the frequency is estimated by fitting the QPO from a PDS which is the average of 16 one-second PDS before and 16 one-second PDS after the time bin (to avoid any possible biases, the PDS of the central time bin is excluded in the average). Then each 1 s PDS is shifted in frequency and averaged to produce one single PDS from which the QPO can be fitted to obtain \dnsaus. We have checked through simulations of a pulsar-like signal, drifting in frequency by about 4 Hz per 32 s (to reproduce the width measured in segment 2, Table \ref{DBarret_t1}), that the method described above is accurate and able to recover the pulsar nature of the signal (in the shifted PDS, the power is concentrated into a single frequency bin). 

\begin{figure} 
\includegraphics[width=0.45\textwidth]{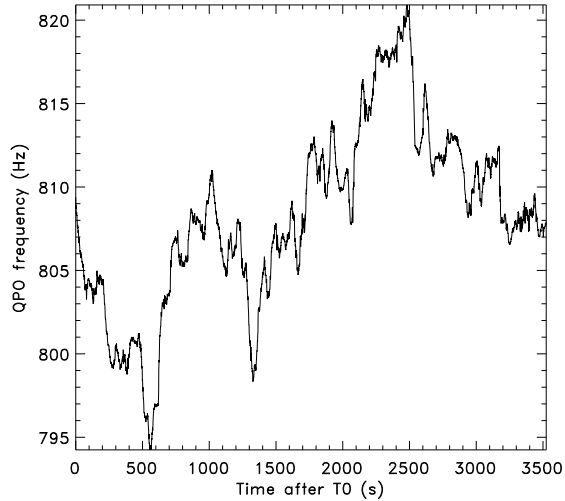}
\caption[]{Time evolution of the QPO frequency in segment 5 as followed every 1 second using a sliding window of 32 s.}
\label{DBarret_f3}
\end{figure}

For maximum accuracy, we have applied this method to the segments of data in which the QPO is almost always (more than 90\% of the time) detected in 32 s time intervals (segments 2-6, 13-18). An example of the time evolution of the QPO frequency measured every second using the sliding window technique is shown in Figure \ref{DBarret_f3}. The results of the fits to the shifted and added 1 s PDS are listed in Table \ref{DBarret_t2}, together with \nudrift\ and $\overline{\Delta \nu_{\rm 1s}}$ from equation \ref{DBarret_eq1}. As can be seen, the recovered \dn\ are systematically larger than would be expected from equation \ref{DBarret_eq1}, and are actually larger than or equal to those obtained by shifting and adding the 32 s PDS. 

There are two possible effects that may combine to explain this quite surprising result. First, the individual frequency estimates have typical error bars (between 0.5 and 1 Hz) of the order of the gain expected in the shift-and-add process. These errors will naturally introduce a blurring of the QPO profile in the shifted PDS. We have checked this effect using simulations of a QPO signal of known frequency, and similar amplitude and width as the real data. Alternatively, this might be an indication that the phase, amplitude, or frequency of the underlying QPO signal vary on time scales significantly shorter than 32 s. 

 \begin{table}
\caption{The results of shifting and adding the PDS on a 1 s timescale for those segments of data where the QPO is detected in more than 90\% of the 32 s time intervals. \dnb=\dnsatd is the same as Table 1. \nudrift\ is the mean absolute frequency drift between consecutive 32 s averaged PDS, $\overline{\Delta \nu_{\rm 1s}}$ is the width expected assuming a linear drift within the 32 s time interval, and \dnsaus\ is the fitted width of the QPO in the shifted and added PDS. }
\begin{center}
\begin{tabular}{ccccccccccc}
Seg & \dnb=\dnsatd & \nudrift & $\overline{\Delta \nu_{\rm 1s}}$ & \dnsaus \\
\hline
\hline
 2 &  4.1$\pm$ 0.1 &  1.3 & 3.9 &  4.2$\pm$ 0.1 \\
 3 &  4.2$\pm$ 0.1 &  1.2 & 4.0 &  4.3$\pm$ 0.1 \\
 4 &  4.2$\pm$ 0.1 &  1.4 & 4.0 &  4.2$\pm$ 0.1 \\
 5 &  3.8$\pm$ 0.1 &  1.4 & 3.5 &  3.9$\pm$ 0.1 \\
 6 &  4.0$\pm$ 0.1 &  1.4 & 3.7 &  4.0$\pm$ 0.1 \\
 7 &  4.9$\pm$ 0.1 &  1.7 & 4.7 &  5.2$\pm$ 0.1 \\
13 &  4.3$\pm$ 0.2 &  1.5 & 4.1 &  4.7$\pm$ 0.2 \\
14 &  4.2$\pm$ 0.1 &  1.6 & 3.9 &  4.5$\pm$ 0.1 \\
15 &  4.3$\pm$ 0.2 &  1.6 & 4.0 &  4.4$\pm$ 0.2 \\
16 &  4.7$\pm$ 0.2 &  2.2 & 4.2 &  5.2$\pm$ 0.2 \\
17 &  4.2$\pm$ 0.1 &  1.9 & 3.7 &  4.4$\pm$ 0.2 \\
\hline
\end{tabular}
\end{center}
\label{DBarret_t2}
\end{table}

\subsection{Coherence time of the oscillator}
In this paper we define the coherence time of the oscillator on the assumption that the signal consists of sine wave shots with exponentially decaying amplitudes with a time constant $\tau$. Such a signal will produce a Lorentzian with a FWHM \dn=$1/(\pi\tau)$ in the PDS.

\begin{figure} 
\includegraphics[width=0.45\textwidth]{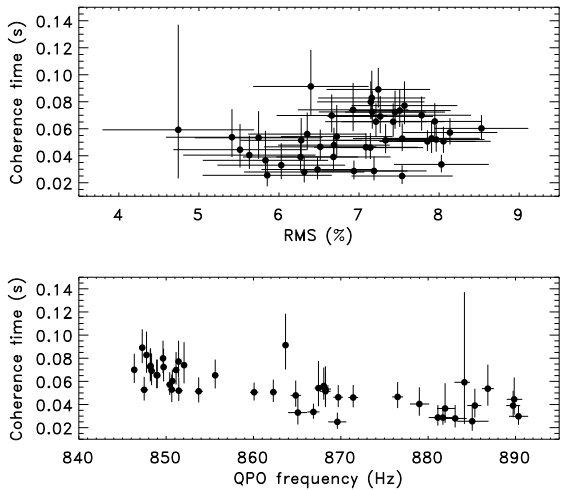}
\caption[]{Top) QPO coherence time as a function of QPO RMS for segment 1, when the
QPO disappeared at about 890 Hz. (Bottom) QPO Coherence time versus QPO frequency for the same segment of data. To reduce scattering in the data points, the averaging time for the PDS is 64 seconds instead of 32 s.}
\label{DBarret_f4}
\end{figure}

Berger et al. (1996) reported a correlation between \dnu~and the frequency, as well as an anticorrelation between the RMS and frequency, using data from the first segment analysed here and a PDS integration time of 100 seconds. As can be seen in Figure \ref{DBarret_f1}, this segment of data is remarkable because the QPO reaches its highest frequency and seems to weaken as it does so. In Figure \ref{DBarret_f4}, we show the dependence of the inferred coherence time versus QPO RMS and QPO frequency. There is a strong anti-correlation between the frequency and coherence time (Spearman correlation coefficient of -0.75 corresponding to a null-hypothesis probability of $1.9~10^{-9}$), and a weaker correlation between the coherence time and RMS (Spearman correlation coefficient of 0.28, corresponding to a null-hypothesis probability of $0.06$). 

\begin{figure} \includegraphics[width=0.45\textwidth]{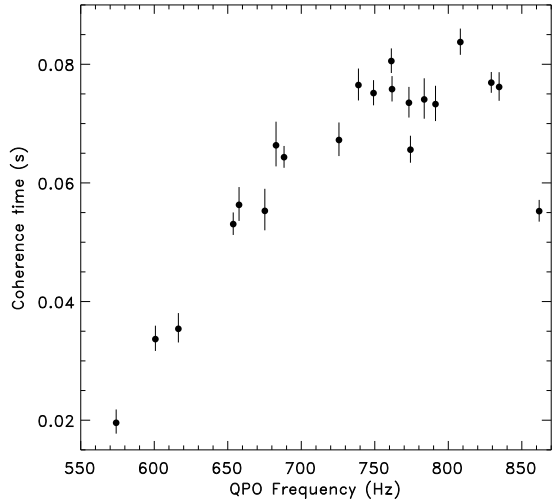}
\caption[]{Dependence of the QPO coherence time on frequency.}
\label{DBarret_f5}
\end{figure}

We have investigated whether these trends were present in the
whole data set, in which the frequency spans a much
wider range. The relationship between the coherence time and \nuqpo\ for
all the data analysed here is shown in Figure \ref{DBarret_f5}. To compute the coherence time, we have used \dnb\ from Table \ref{DBarret_t1}. There is a clear
{\it positive} correlation between the two quantities until \nuqpo\ 
reaches a maximum level around $\sim 800$ Hz (albeit with some scatter). 
After the maximum, an anticorrelation between the QPO frequency and coherence time, of the type seen within
segment 1, seems to be present. A similar behavior was reported from 4U1636-53 for its lower kilo-Hz QPO (di Salvo et al. 2003; see also van der Klis et al. (1997) for the upper kilo-Hz QPO of Sco X-1).

\begin{figure} \includegraphics[width=0.45\textwidth]{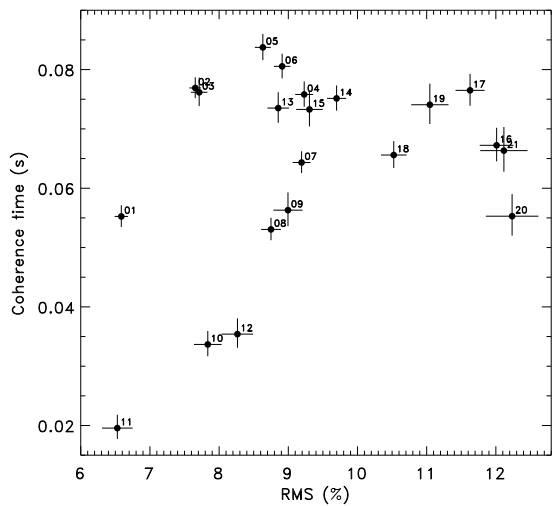}
\caption[]{Dependence of the QPO coherence time on RMS. The numbering refers to Table \ref{DBarret_t1}.}
\label{DBarret_f6}
\end{figure}

The overall relationship between the RMS amplitude of the QPO and its coherence time is shown in
Figure \ref{DBarret_f6}. The behaviour is much more complex; both positive and
negative anticorrelations are seen. Again, the pattern seen in
segment 1 is not typical.

\begin{table*}
\caption{The higher-frequency QPOs detected after applying the
shift-and-add procedure to the 32 s PDS. $\nu_{\rm shift}$ is the frequency to which the 32 s PDS were shifted. $\Delta\nu_{\rm low}$ is the width of the low-frequency QPO. $\nu_{\rm high}$, $\Delta\nu_{\rm high}$ and RMS$_{\rm high}$ are respectively the frequency, the width and the RMS of the high-frequency QPO. To obtain reliable fits, segments 1 to 3, 13 to 15, 16 to 18 and 20-21 were grouped together. The errors are again computed as $\chi^2=\chi^2_{\rm min}+1$.}
\begin{center}
\begin{tabular}{lcclll}
Seg & $\nu_{\rm shift}$ & $\Delta\nu_{\rm low}$ & $\nu_{\rm high}$ & $\Delta\nu_{\rm high}$ & RMS$_{\rm high}$ (\%) \\
\hline
 1-3 &  843.5$^{+0.03}_{-0.03}$ &    4.5$^{+ 0.1}_{- 0.1}$ & 1103.9$^{+17.9}_{-17.1}$ & 179.6$^{+142.7}_{- 70.2}$ &  4.1$\pm$0.2\\
 7 &  688.5$^{+0.05}_{-0.05}$ &    4.9$^{+ 0.1}_{- 0.1}$ &  994.0$^{+ 3.1}_{- 3.0}$ &  26.1$^{+ 15.3}_{-  9.7}$ &  3.4$\pm$0.7\\
 8 &  653.4$^{+0.08}_{-0.08}$ &    6.0$^{+ 0.2}_{- 0.2}$ &  966.7$^{+ 1.2}_{- 1.8}$ &  11.7$^{+  7.9}_{-  8.6}$ &  2.9$\pm$0.8\\
10 &  600.5$^{+0.20}_{-0.20}$ &    9.5$^{+ 0.6}_{- 0.6}$ &  901.8$^{+ 8.2}_{- 7.0}$ &  87.4$^{+ 44.1}_{- 35.5}$ &  6.4$\pm$0.5\\
11 &  570.1$^{+0.55}_{-0.55}$ &   16.3$^{+ 1.7}_{- 1.7}$ &  879.5$^{+ 4.1}_{- 3.9}$ &  89.8$^{+ 20.6}_{- 16.5}$ &  8.4$\pm$0.4\\
12 &  611.1$^{+0.21}_{-0.21}$ &    9.0$^{+ 0.6}_{- 0.6}$ &  930.7$^{+ 8.8}_{- 7.7}$ & 101.1$^{+ 46.5}_{- 35.4}$ &  8.0$\pm$0.5\\
13-15 &  771.5$^{+0.03}_{-0.03}$ &    4.3$^{+ 0.1}_{- 0.1}$ & 1037.7$^{+14.4}_{-12.9}$ & 134.6$^{+ 83.4}_{- 51.6}$ &  5.1$\pm$0.4\\
16-18 &  745.5$^{+0.03}_{-0.03}$ &    4.6$^{+ 0.1}_{- 0.1}$ & 1039.0$^{+13.8}_{-13.1}$ & 115.6$^{+ 76.6}_{- 51.8}$ &  5.7$\pm$0.5\\
20-21 &  677.6$^{+0.08}_{-0.08}$ &    5.3$^{+ 0.2}_{- 0.2}$ &  985.1$^{+14.9}_{-15.0}$ & 182.1$^{+176.1}_{- 75.6}$ & 11.0$\pm$0.7\\
\hline
\end{tabular}
\end{center}
\label{DBarret_t3}
\end{table*}

\subsection{Coherence of the higher-frequency QPO}
We have searched for the second QPO in the data using the shifted 32 s PDS
used above.  We detect an otherwise invisible higher-frequency QPO in 5 of the
21 segments used here. A weaker signal is detected in segments 1 to 3, 13 to 15, 16 to 18 and 20-21, which we group to obtain a meaningful fit.  The results of fitting the higher-frequency
QPOs are listed in Table \ref{DBarret_t3}. Given the relatively large width of the high-frequency QPO, the window for the fit is 200 Hz on each side of the QPO peak, excluding the region around the lower-frequency QPO. These QPOs have been previously
reported in \citet{mendez98a,mendez98b,mendez99}. What is noticeable is that the higher-frequency QPO is much broader
than the lower one, with Q less than $\sim 10$. This is a well known observational fact
\citep{vdk2}. Interestingly, there is no apparent correlation between the
coherence time of the higher frequency QPOs and its frequency (of
the type seen in Figure \ref{DBarret_f5} for the lower-frequency QPO), but we note
that the frequencies do not span a wide range and that the error bars
are relatively large.

\section{Discussion}
 We have carried out a systematic study of the kilo-Hz QPO of 4U1608-52, with the aim of investigating  the coherence time of the underlying oscillator. By following the lower kilo-Hz QPO over the shortest timescales permitted by the data statistics (down to 32 s) to minimise the contribution of the long-term frequency drifts, with a shift-and-add technique, we have found a lower limit on the quality factor larger than $\sim 150$ in 14 of the 21 segments of analysed data. We have shown that in this object Q is as high as 200 for thousands of seconds. In many sources, the strength of the signal does not allow the study of the QPOs on such short timescales, which may explain why the Q reported in the literature are generally lower than this.

Frequency drifts as large as 5 Hz are observed between consecutive 32 s PDS. In an attempt to remove the contribution of this short-term drift to the width of the QPO, we have applied the shift-and-add procedure to the 1 s PDS, using a sliding-window technique to estimate the QPO frequency every second. If the frequency drift were linear over 32 s, this method should remove the drift contribution to the width measured on 32 s and provide an estimate of the QPO width on a 1 s timescale. Interestingly enough, the recovered QPO width is not significantly smaller than the width measured on 32 s. This surprising result could be explained by the fact that the error bars on the frequency are of the order of the expected gain, causing an artificial blurring of the QPO profile in the shift-and-add process. Alternatively, this might be indication that the frequency, amplitude, or phase of the underlying signal varies on time scales much shorter than 32 s. In this case, this would most likely imply that the Q factor reported in Table \ref{DBarret_t1} with the present technique underestimates the intrinsic Q of the QPO.

Determining the coherence time of the underlying oscillator from the width of the QPO over the wide range of frequency spanned (between 560 Hz and 890 Hz), we have shown that there is a clear pattern in which the coherence time increases with frequency up to a maximum at $\sim 800$ Hz. At both ends of the frequency range, the decreasing strength of the QPO seems to be associated with a decrease of {\it both} its
amplitude and its coherence.

We now discuss the implications of the above results for QPO
models. There are two different  classes of models of the kHz
QPOs: those involving clumps within or above the accretion disk
(e.g., \cite{miller98,stella99},\cite{vdk3} for a recent review), and those
involving oscillations of the disk (reviewed in \cite{wagoner99}, and
 \cite{kato01}).

\subsection{Clumps}
If inhomogeneities can form in the accretion flow in the form of
clumps  which are more luminous than their surroundings and which
orbit around the central star, the viewing geometry might be such
that the clumps will produce luminosity variations \citep{bath73}.
Inhomogeneities lying at a radius R and having a radial extent
$\Delta R$ will be sheared by differential rotation and will
therefore have a finite lifetime. The maximum lifetime for such a
clump to be stretched to an axisymmetric ring is given by $\tau_s= (2/3)(R/\Delta R) P_k$, where $P_k$ is
the Keplerian period at radius R \citep{bath74}. In the  clump model, this lifetime
         is comparable to the  coherence time of the QPO $\tau \sim \tau_s $. Such clumps will
produce a luminosity variation of at most $\Delta L/L=\Delta R/R$
\citep{pringle81}. Hence, $\tau_s \propto R/\Delta R \propto (\Delta L/L)^{-1}$,
           so high coherence of the signal and its high amplitude
           are incompatible in the clump model. In our case $\Delta L/L$ is of the order of
10\% (see Figure \ref{DBarret_f6}). This would imply $\tau_s \sim 10 P_k \la 0.01$
second (or Q $\sim 30$), much shorter than the coherence time we
found in our analysis (see Figure \ref{DBarret_f5}). It is also worth noting that in the above formula the
clump lifetime (and signal amplitude) are predicted to decrease with increasing frequency, (e.g. Livio \& Bath 1982), unlike the observed behaviour in Figures \ref{DBarret_f5} and \ref{DBarret_f6}.


This is clearly not what we observe in this source (see
Figure \ref{DBarret_f6}). Alternatively, one can make the assumption that the
radial extent the blob is of the order of the disk height ({\it h}). Shakura and Sunyaev (1973)  estimate
     $h \sim 3 \times 10^6 {\rm cm} (L/L_{\rm Edd}) (M/M\odot)$,
     and the radius of the relevant orbit is on the order of the stellar 
     radius, or a few times larger.
For an accreting neutron star with a large fraction of Eddington
luminosity, $h/R\sim1/10$ is a reasonable estimate, which again
leads to the same conclusions. 

We have also found that the higher-frequency kilo-Hz QPO is much
fainter  and is characterised by a much lower quality factor. This
result is hard to reconcile with  models  in which the higher-frequency QPO is a Keplerian frequency at some radius (e.g., the
sonic point, \cite{miller98}), and the lower QPO is a beat
frequency generated by interaction of the neutron star radiation
with clumps originating at that particular radius. This model
predicts that the width of the two QPOs should be similar, because
the stellar spin frequency is expected to be highly coherent. In
4U1608-52, we have shown the width of the higher-frequency QPO to
be much larger than the width of the lower-frequency QPO.

Another model of QPOs has been proposed in which clumps leave the
accretion  disk and for some reason follow test particle orbits.
In this so-called relativistic precession model (e.g. Stella \& Vietri 1999), the clump is
responsible for all three frequencies: two kilo-Hz QPOs and a
lower-frequency QPO, also called the horizontal branch oscillation. Since the same agent is responsible for the two
kHz QPOs, the coherence times should be about the same. We have
shown here that the coherence time of the higher-frequency QPO is
at least a factor of ten lower than that of the lower-frequency
QPO (See Table 3).

More stable non-linear structures (vortices) may persist in the
accretion  disk, but these will be subject to radial inflow unless
they are anchored by an external magnetic field
\citep{abramowicz92,lovelace99,tagger01}. The inflow velocity at
radius $r$ in an  $\alpha$-disk of thickness $h$, is a factor of
$\alpha(h/r)^2$ lower than the orbital velocity, where
$\alpha>10^{-2}$ is the dimensionless viscosity. So, for typical
disks, within $\sim 10$ orbits the fluid is carried inwards by a
distance equal to a few per cent of its initial radius, and
suffers a percentage change in orbital frequency which is 1.5
times larger.

Thus all models of this class seem to be incompatible with a high Q factor, and the relationships found between the coherence time with frequency and RMS amplitudes.

\subsection{Disk oscillations}
There is growing evidence that the oscillations originate in the
disk  itself, in particular because the same phenomena are
observed in a wide range of accreting X-ray sources, from white
dwarfs to black holes \citep{mauche02}. Several models, relating
QPOs to disk oscillations have been proposed, some of them making
clear predictions of what the quality factor of the QPO should be.
This is the case, for instance, for the transition layer model
proposed by \cite{tita98}. In this model, a shock occurs where the
Keplerian disk adjusts to the sub-Keplerian flow and the
transition can undergo various types of oscillations under the
influence of the gas, radiation, magnetic pressure, and
gravitational force. In this case Q can in principle reach 100. It
is expected to increase with frequency but also with luminosity
(\cite{tita98}). In our data set it is clear that there is not a
good correlation between Q and the source count rate.

For acoustic waves in isotropic turbulence \cite{goldreich88}
predict $Q\sim h/(\lambda\alpha)$. Here $\alpha\sim0.1$ is
viscosity parameter for a Shakura-Sunyaev disk and the wavelength,
$\lambda$, is much greater than $h$,  yielding a rather low value
of $Q$. Anisotropic turbulence increases this to $Q=\lambda
/(h\alpha) \sim10$, as observed in black holes. For g-modes and
isotropic turbulence, $Q$ is the higher value  of $1/\alpha$ and
$\sqrt{r/h}$, but the highest values of Q are obtained for
$g$-modes in anisotropic turbulence, $(r/h)/\alpha\sim 100$. For
c-modes, which correspond to a warp revolving at  the
Lense-Thirring frequency (a general-relativistic effect), this
value is enhanced by a factor approximately equal to the square of
the ratio of orbital velocity to the speed of sound
(the discussion in this section follows \citet{ortega00}). In the very hot disk around the neutron stars
this enhancement may still be insufficient to explain the high
Q reported here. However, some modes are excited, rather
than damped, by viscosity and may enter the non-linear regime
\citep{ortega00}.

\section{Conclusions}
We have shown that the kilo-Hz QPO in 4U1608-52 is a highly
coherent signal with an inferred coherence time of up to about one
tenth of a second. This result, as well as the dependence of the
QPO coherence time on frequency and amplitude which we report are hard to reconcile with
QPO models involving orbiting clumps. This leaves the accretion
disk oscillations as the the most likely alternative for accounting
for these signals. 
\section*{Acknowledgments}
We are grateful to Mariano M{\' e}ndez, Michiel van der Klis and Bob Wagoner for
very helpful discussions. WK acknowledges support by CNRS through a {\sl poste rouge}, and thanks CESR for hospitality.
Research supported in part by LEA Astro-PF, and KBN through grant 2 P03D 014 24. SP acknowledges a grant from the Swiss National Science Foundation. 

We are grateful to an anonymous referee whose detailed and comprehensive comments helped to improve the clarity of the points made in this paper.

\end{document}